# Molecular Dynamics Simulation of Thermal Boundary Conductance Between Carbon Nanotubes and SiO$_2$

Zhun-Yong Ong[1,2] and Eric Pop[1,3,4,*]

[1]*Micro and Nanotechnology Lab, Univ. Illinois at Urbana-Champaign, Urbana IL 61801, U.S.A.*
[2]*Dept. of Physics, Univ. Illinois at Urbana-Champaign, Urbana IL 61801, USA*
[3]*Dept. of Electrical & Computer Engineering, Univ. Illinois at Urbana-Champaign, Urbana IL 61801, USA*
[4]*Beckman Institute, Univ. Illinois at Urbana-Champaign, Urbana IL 61801, USA*

We investigate thermal energy coupling between carbon nanotubes (CNTs) and SiO$_2$ with non-equilibrium molecular dynamics simulations. The phonon thermal boundary conductance ($g$) per unit length is found to scale proportionally with the strength of the Van der Waals interaction (~$\chi$), with CNT diameter (~$D$), and as power law of temperature (~$T^{1/3}$ between 200–600 K). The thermal relaxation time of a single CNT on SiO$_2$ is independent of diameter, $\tau \approx 85$ ps. With the standard set of parameters $g \approx 0.1$ WK$^{-1}$m$^{-1}$ for a 1.7 nm diameter CNT at room temperature. Our results are comparable to, and explain the range of experimental values for CNT-SiO$_2$ thermal coupling from variations in diameter, temperature, or details of the surface interaction strength.

## I. INTRODUCTION

Energy transport across the interfaces of nanostructures and their surrounding medium plays a critical role in the performance and stability of carbon nanotubes (CNTs), graphene nanoribbons, or other nanomaterials. For instance, thermal transport through CNT composites is dominated by the small interface thermal conductance,[1,2] rather than the large intrinsic thermal conductivity of the nanotubes themselves.[3-5] The thermal conductance of CNT arrays with potential heat sinking applications is similarly restricted at the interface with the silicon chip substrate (Fig. 1a).[6,7] In CNT transistors and interconnects, the maximum current-carrying ability of the devices is limited by self-heating and heat dissipation with the dielectric substrate (Fig. 1b).[8,9] In all these cases, energy exchange from the CNTs to the environment is determined by the thermal boundary conductance (TBC) at their interface. In particular, the nanoscale constriction between the CNT and the substrate or environment, combined with the weak bonds at this interface are expected to lead to a small TBC.[10,11]

In this context, it remains highly desirable to examine the atomistic details of the nanotube-substrate thermal coupling in order to understand the different channels of heat dissipation. Moreover, the dependence of TBC on nanotube diameter, temperature, and Van der Waals interaction strength are presently unknown. In addition, the predominant dielectric within integrated circuits is $SiO_2$, which also forms a thin surface layer (native oxide) on the backside of silicon chips, where CNT heat sinks would be attached. Thus, the primary goal of this paper is to investigate the lattice (phonon) contribution to the TBC between single-walled carbon nanotubes and amorphous $SiO_2$ using the molecular dynamics (MD) simulation technique. The MD technique has been previously used to examine the TBC between nanotubes and surrounding liquid or solid media.[12-16] Here, we study the TBC with $SiO_2$ as a function of the substrate temperature as well as the interaction strength between the substrate and the CNT. We also examine the dependence of the TBC on diameter in armchair nanotubes. Finally, we compare our results with existing thermometry data for the TBC of CNTs on $SiO_2$ substrates, and provide physical insight into the observed trends.[17-19]

## II. MOLECULAR DYNAMICS SETUP

The molecular dynamics simulation technique treats the atoms in the system classically.[20] The individual atoms (here C, O, and Si) are modeled as point masses interacting through a given set of interatomic potentials, as detailed below. The atomic equations of motion are integrated numerically to produce the classical trajectories of the system. To perform our simulations, we use the molecular dynamics code LAMMPS[21] because of its parallelization and the implementation of the interatomic potentials used. Our simulations were run on a Linux cluster which consists of eight 3.0 GHz processor cores (Intel 5400 series) with 16 GB of memory.

**A. Description of CNT and SiO$_2$**

To model the interactions between carbon atoms (C-C) in the molecular dynamics simulation, we use the adaptive intermolecular reactive empirical bond order (AIREBO) potential[22] derived from the second generation Brenner potential.[23] The AIREBO potential is widely used in MD simulation of CNTs and has been implemented in LAMMPS. The AIREBO potential also reproduces the phonon density of states (DOS) accurately, with a cutoff at approximately 56 THz, as shown in Fig. 1c.

To model the Si-Si, Si-O and O-O atomic interactions, we use the recently published Tersoff-type potential parameterization of Si-O systems by Munetoh,[24] which we will henceforth refer to as the Munetoh potential. The Munetoh potential reproduces the vibrational DOS of bulk amorphous SiO$_2$ in good agreement with known experimental data.[25] The structure of the amorphous SiO$_2$ produced with the Munetoh potential has been shown to be that of a three-dimensional random network of SiO$_4$ tetrahedral units.[24, 26] The vibrational DOS for the Si and O atoms in the substrate are also shown in Fig. 1c, noting that Si and O phonon spectra do not extend beyond 40 THz. Although our choice of interatomic potentials for SiO$_2$ does not take into account long-range Coulombic interaction, this is less likely to matter for the problem at hand as the interfacial thermal transport depends only on the Van der Waals (VdW) interaction between the CNT and substrate atoms. The main consideration in our choice of the interatomic potentials for the nanotube and the substrate is the realism of the vibrational DOS of the simulated atoms.

**B. Interaction between CNT-SiO$_2$**

There is some uncertainty in the strength and form of the interactions between the CNT atoms and the substrate atoms. It has been proposed that the interaction between the CNT and substrate is primarily short-range Van der Waals.[27, 28] Thus, the Si-C and O-C interactions are modeled as VdW interactions using the 12-6 Lennard-Jones (LJ) potential function, written as

$$V(r) = 4\chi\varepsilon\left[\left(\frac{\sigma}{r}\right)^{12} - \left(\frac{\sigma}{r}\right)^{6}\right] \tag{1}$$

where $\varepsilon$ is the energy parameter, $\sigma$ the distance parameter, $r$ the interatomic distance and $\chi$ is a scaling factor. The $\varepsilon$ parameters determine the strength of the specific interactions between the nanotube and the substrate atoms. We have two $\varepsilon$ parameters ($\varepsilon_{Si-C}$ and $\varepsilon_{O-C}$) and two $\sigma$ parameters ($\sigma_{Si-C}$ and $\sigma_{O-C}$). The general scaling factor $\chi$ adjusts the overall nanotube-substrate interaction. The parameters used in our simulations are based on those for VdW interactions in the Universal Force Field (UFF) model by Rappe et al.[29] The UFF-based parameters are $\varepsilon_{Si-C}$ = 8.909 meV, $\varepsilon_{O-C}$ = 3.442 meV, $\sigma_{Si-C}$ = 3.326 Å and $\sigma_{O-C}$ = 3.001 Å. The parameters are calculated using the Lorentz-Berthelot mixing rules.[20] The cut-off distances of the LJ potential for the Si-C and O-C interactions are set equal to 2.5$\sigma$, or 8.315 Å and 7.503 Å respectively.[30] To study the effect of the substrate-CNT interaction strength, we repeat our simulations with different values of $\chi$. We use $\chi$ = 1, 2 and 4 for the scaling factor in the simulations, with $\chi$ = 1 corresponding to the original form of the LJ interactions as used in the UFF.

**C. Preparation of Simulation Domain**

The nanotube-substrate domain is prepared as shown in Fig. 2. It is relevant to point out we focus on the "horizontal" nanotube-SiO$_2$ interaction, as opposed to the vertical case which has been previously studied for nanotube-Si thermal transfer.[15, 16] We believe the horizontal scenario is more relevant to heat dissipation from CNTs in both heat sink applications as well as interconnects, as shown in the schematics of Fig. 1a and 1b. Even for vertical CNT arrays with potential heat sink applications,[6, 7] the CNTs contacting the substrate will most likely do so at a "bent" angle, more consistent with the lateral coupling studied here. Moreover, silicon wafer substrates are typically covered with a thin, native SiO$_2$ layer (1-2 nm thick), such that the relevant interfacial atomic interaction is that between the CNTs in the array and the surface SiO$_2$.

To build the amorphous SiO$_2$ layer we first replicate a β-cristobalite unit cell within a rectangular simulation domain of 57.3 × 28.7 × 36.9 Å with periodic boundary conditions in all three directions (Fig. 2a). To obtain the amorphous bulk SiO$_2$, we anneal the crystalline SiO$_2$ atoms at the temperature of 6000 K for 10 ps and at fixed pressure of 1 bar using a time step of 0.1 fs (Fig. 2b). Then, we slowly quench the structure to 300 K at a rate of $10^{12}$ K/s. To create the SiO$_2$ surface from the amorphous structure, the silica atoms in the top half of the simulation do-main are deleted to obtain an amorphous SiO$_2$ slab with 3904 atoms. Using the conjugate gradient algorithm, we perform an energy minimization of the resultant structure to produce the final SiO$_2$ surface. A 36.9 Å long, 600-atom (10,10) CNT was constructed and inserted into the top half of the simulation domain just above the amorphous SiO$_2$ slab. Once again, we perform an energy minimization to obtain the final substrate-CNT structure shown in Fig. 2c.

### III. SIMULATION OF CNT-SUBSTRATE THERMAL COUPLING

The phonon density of states (DOS) of the carbon nanotube and SiO$_2$ substrate are proportional to the Fourier transform of the velocity autocorrelation function (VACF). The normalized velocity autocorrelation function (VACF) of an atom can be written as

$$f(t) = \frac{\langle \vec{v}(0) \cdot \vec{v}(t) \rangle}{\langle \vec{v}(0) \cdot \vec{v}(0) \rangle} \tag{2}$$

where $\vec{v}(t)$ is the velocity of the particle at time $t$ and the angled brackets $\langle \ldots \rangle$ represent the ensemble averages. For long simulation times, the ensemble averages can be replaced by the time averages. The calculated phonon (vibrational) DOS of the C atoms in the CNT and the Si and O atoms in the substrate are shown in Fig. 1c and Fig. 6c.

Once the density of states is known it is necessary to define a local temperature $T$. We use the 'kinetic' definition for an atomistic system as

$$T = \frac{1}{3Nk_B} \sum_{i=1}^{N} m_i v_i^2 \tag{3}$$

where $N$ is the number of atoms in the system, $m_i$ the mass of the $i$-th atom and $v_i$ its velocity. In order to simulate the interfacial heat transfer, we have to set up an initial temperature difference $\Delta T$ between the CNT and substrate atoms, with the CNT at the higher temperature. In the absence of additional coupling to an external heat reservoir, $\Delta T$ decays exponentially with a single relaxation time ($\tau$) such that as

$$\Delta T(t) = \Delta T(0) e^{-t/\tau}. \tag{4}$$

Given that the thermal resistance of the nanotube-substrate interface is much greater than the internal thermal resistance of the nanotube, we can apply a lumped heat capacity method, as sche-matically shown in the Fig. 2c inset. Thus, the thermal boundary conductance (TBC) per unit nanotube length ($g$) is given by

$$g = \frac{C_{CNT}}{\tau} \qquad (5)$$

where $C_{CNT}$ is the heat capacity per unit length of the nanotube. We use the definition of TBC per unit length rather than the more conventional conductance per unit area[31] because the contact area between the CNT and substrate is not easily defined. However, the TBC per unit area could be approximated by normalizing via the CNT diameter, as we do later in Fig. 6 and surrounding discussion. In our simulations, we set the CNT-substrate temperature difference $\Delta T$ to be a fraction of the initial substrate temperature. If $\Delta T$ is too small with respect to the substrate temperature, the relaxation process will be very noisy. On the other hand, if $\Delta T$ is too large with respect to the substrate temperature, it becomes difficult to determine the temperature dependence of the TBC. As a compromise, we set $\Delta T$ to be one half of the initial substrate temperature.

In order to produce the temperature difference between the CNT and the substrate, we equilibrate the atoms in the CNT and the top four-fifths of the slab in Fig. 2c at the desired temperature $T$ for 100 ps. The atoms in the bottom fifth of the slab are always kept frozen (motionless) to anchor the substrate. The atoms are set to a given temperature using velocity rescaling with a time step of 0.25 fs. Afterwards, the velocity rescaling algorithm is switched off and the system is allowed to equilibrate. To produce a temperature difference between the CNT and substrate, we again apply the velocity rescaling algorithm to the substrate atoms at the temperature $T$ and to the CNT atoms at temperature $T+\Delta T$ for 10 ps.

To simulate the heat transfer process, the velocity rescaling is switched off and the system is allowed to relax, producing temperature transients. We then record the decay of the temperature difference between the nanotube and SiO$_2$, as shown in Fig. 2d. Because this temperature decay $\Delta T$ is noisy, it is necessary to average over multiple runs (ten in our case) to obtain the exponential decay behavior shown in Figs. 2d, 3a and 3b, and expected through the lumped model from Eq. (5). Throughout this work, we repeat such simulations for several VdW strength scaling factors ($\chi$ = 1, 2 and 4), as shown in Figs. 3a and 3b. Moreover, the cross-sectional profile of a (10,10) nanotube on SiO$_2$ is shown for the three interaction strengths in Fig. 3c. As expected, the CNT profile becomes progressively deformed as the interaction strength increases. The mini-mum CNT-SiO$_2$ equilibrium separation is found to be approximately 2.2 Å, in relative agreement with a recent density functional theory (DFT) study of the CNT-quartz interaction,[32] which found a separation of 2.5 Å.

## IV. SIMULATION RESULTS

### A. Temperature Dependence of TBC

We perform our simulations at different substrate temperatures ($T$ = 200, 300, 400, 500 and 600 K) with different VdW strength scaling factors ($\chi$ = 1, 2 and 4). Figure 4a shows our calculations of the thermal boundary conductance for the (10,10) nanotube interface with the amorphous SiO$_2$ substrate. The results show that the TBC increases monotonically both with temperature and with VdW coupling strength $\chi$. The temperature dependence of the TBC for the horizontal CNT is consistent with the temperature dependence of TBC for vertically aligned CNTs[16] with Si substrates, and other previous simulations.[33, 34] In the classical limit, this proportionality with temperature suggests that inelastic scattering of phonons at the interface plays a significant role in the thermal relaxation of the CNT.[34] By contrast, continuum models of thermal interfaces such as the acoustic mismatch model (AMM) or the diffusive mismatch model (DMM)[35] only assume elastic scattering; i.e. for each incident phonon, a single phonon with the same frequency will scatter from the interface. The MD simulation technique intrinsically accounts for inelastic scattering, such that phonons can break down into several lower frequency modes, or conversely multiple phonons can scatter into higher frequency modes. Stevens et al[34] similarly reported a linear dependence of the TBC on temperature in their MD simulation of solid-solid interfaces using the LJ potential, and attributed the temperature dependence to the inelastic scattering of phonons at the interface. In addition, this monotonic rise of TBC with temperature has also been documented experimentally from numerous measurements between dissimilar materials.[36]

To further understand the dependence of the TBC on temperature, our results at different temperatures normalized by the TBC at 200 K are plotted in Fig. 4b. We find that the temperature dependence is remarkably similar for all values of interaction strength ($\chi$) considered. For example, the TBC at 600 K is roughly 40 percent greater than the TBC at 200 K. This implies that the TBC dependence on temperature is independent of the VdW interaction strength ($\chi$) and suggests a simple power law behavior. The data are in good agreement with a $T^{1/3}$ fit over this temperature range, as shown in Fig. 4b.

## B. Interaction Strength Dependence of TBC

While the *scaling* of TBC with temperature appears independent of the CNT-substrate interaction strength ($\chi$), the magnitude of the TBC itself is expected to depend on $\chi$. To determine the effect of the CNT-substrate interaction strength on the thermal boundary conductance, we replot the TBC results rescaled by $1/\chi$ in Fig. 4c. These suggest that the TBC is directly proportional to the substrate-CNT interaction strength, in contrast to the conventional diffuse and acoustic mismatch models (DMM and AMM)[35] which do not capture the atomistic interaction and atomic arrangement at the interface. It should also be noted that the DMM and AMM are usually applied to dissimilar *bulk* materials and are not necessarily expected to produce good agreement for nanomaterial interfaces. Thus, atomistic molecular dynamics investigations, such as the present work, are essential in exploring and explaining heat and energy transfer across the interfaces of carbon nanotubes and other nanomaterials.[12-16]

The linear dependence of the TBC on $\chi$ can be better understood by considering the interatomic coupling of CNT and substrate atoms. For illustrative purposes, we consider only the harmonic limit. The energy of the *p*-th atom can be written as

$$E_p = \frac{1}{4}\sum_q \left(u_p^* H_{pq} u_q + u_q^* H_{qp} u_p \right) + \frac{1}{2} M_p \dot{u}_p^* \dot{u}_p \qquad (6)$$

where $u_p$ and $M_p$ are the displacement and mass of the *p*-th atom, and $H_{pq}$ is the second derivative of the interatomic potential i.e.

$$H_{pq} = \frac{\partial^2 \phi}{\partial u_p \partial u_q} \qquad (7)$$

and $\phi$ is the interatomic potential. If we take the time derivative of $E_p$, use Newton's second law

$$M_p \ddot{u}_p = -\sum_q H_{pq} u_q \qquad (8)$$

and sum over the atoms in the CNT, then we obtain following expression

$$\sum_p \frac{dE_p}{dt} = \frac{1}{4}\sum_p \sum_q \left( u_p^* H_{pq} \dot{u}_q + \dot{u}_q^* H_{qp} u_p - u_q^* H_{qp} \dot{u}_p - \dot{u}_p^* H_{pq} u_q \right) \qquad (9)$$

where the sum over $p$ is for the CNT atoms and the sum over $q$ is for the substrate atoms. The $H_{pq}$ terms that survive in Eq. (9) correspond to the second derivative of the LJ interaction between the CNT and the substrate and scale linearly with χ as given in Eq. (1). Physically, Eq. (9) corresponds to the rate of work done on the CNT ($F \cdot v_{CNT}$) minus the rate of work on the substrate (-$F \cdot v_{SiO2}$) by forces ($F$) between the CNT and the substrate atoms. Thus, as we increase χ, which amounts to strengthening the forces between CNT and substrate, the rate of energy dissipation from the CNT atoms to the substrate atoms increases proportionally. As a result, the decay time τ varies inversely with χ, as we have seen in Fig. 3. Therefore the TBC scales proportionally with χ, because increasing the latter also increases the harmonic coupling between the phonons modes in the CNT and the substrate. Although our simple argument was based on harmonic forces, we note that the inclusion of higher order terms in Eq. (6) will not affect the conclusion that the TBC scales linearly with χ; the anharmonic terms between C-C atoms will only affect the forces between CNT atoms, not leading to external dissipation of energy, while the anharmonic terms between CNT-substrate atoms will scale linearly with χ.

The dependence of TBC on the substrate-CNT interaction suggests that a simple strategy for engineering this thermal coupling would be to modify the atomic morphology of the substrate surface. For example, if the surface is passivated with a relatively inert species that does not interact strongly with carbon, the TBC is expected to decrease. Alternatively, if the surface roughness is increased so that the overall physical contact between the nanotube and the substrate is reduced, then the TBC will also decrease. Conversely, by improving the surface smoothness, e.g. along the steps of a miscut crystalline insulating (e.g. quartz) substrate, the TBC should be improved. Similarly, functionalizing nanotubes with polymers or metals that increase both coupling strength, phonon DOS overlap, as well as interaction area should also lead to improved TBC.[37]

**C. Diameter Dependence of TBC**

To study the diameter dependence of the TBC, we repeat our simulations with armchair CNTs of (6,6), (8,8), and (12,12) chirality, corresponding to diameters of 0.81, 1.08 and 1.63 nm, respectively. The setup is identical to our earlier simulation with the (10,10) nanotube except for the size of the system. The cross sections of the CNTs are shown in Fig. 5. The substrate-CNT interaction scaling factor is maintained at χ = 1, and the nanotubes are all approximately 9.84 nm long. In each run, the substrate and the CNT atoms are equilibrated at 300 K and 500 K, respec-tively, for 80 ps using velocity rescaling before the temperature difference between the CNT and the substrate is allowed to decay over 80 ps. We average the temperature decay over five runs for each CNT. As before, the decay time is then obtained by fitting to a single exponential time.

In Fig. 6a, we see that the average thermal relaxation time is approximately constant for the range of diameters used. Since the TBC is equal to the product of heat capacity and the inverse decay time, this implies that the TBC scales linearly with the diameter, as the lattice heat capacity. We plot the TBC vs. CNT diameter in Fig. 6b and observe this linear relationship between them after performing an empirical fit (dashed line). The TBC increases with diameter because a larger diameter means that there are more phonon branches and this implies there are more phonons available to participate in the interfacial thermal transport process. In addition, Fig. 6c shows the normalized phonon DOS for the CNTs of varying diameter. We see that the overall distribution of phonon modes does not change significantly with diameter for armchair CNTs. It is believed that the low-frequency modes are primarily responsible for interfacial thermal transport. Since the *proportion* of low-frequency modes does not change, the absolute number of low-frequency modes available for interfacial thermal transport scales linearly with the CNT diameter. Assuming that the average energy relaxation rate of each phonon mode does not change with diameter, this explains the simple linear scaling of the TBC with diameter over the examined range. A second, more intuitive, explanation is that larger CNTs have more atoms in closer contact with the substrate. These atoms would be more strongly coupled to the substrate and provide more channels of energy dissipation to the substrate.

Another measure of heat dissipation from a nanostructure like the CNT is the TBC divided by the diameter, i.e. the approximate TBC per unit area rather than per unit length. We use this simple convention because a more precise contact area is difficult to define atomistically or to measure in practice. On the other hand, the diameter is a practically accessible quantity, and thus a simpler choice to estimate the "footprint" of a single CNT. Our simulation results suggest that the TBC per unit area with $SiO_2$ does not vary with diameter for armchair CNTs and is approximately $5.8 \times 10^7$ WK$^{-1}$m$^{-2}$ at room temperature. This is a value comparable to that measured for other (bulk) material interfaces, approximately between that for Pb with diamond, and that for Al with $Al_2O_3$.[31, 36] This TBC per unit area is perhaps a more important figure of merit in practical, macroscopic applications such as heat sinks. In this sense, our results indicate that for a given vertical nanotube array packing density (number of nanotubes per unit area) arrays with larger average CNT diameter will benefit from higher thermal conductance at the silicon backside interface, assuming a contact geometry like that depicted in Fig. 1a.

Before concluding, it is also important to compare our simulation results with the few existing experimental studies of the CNT-SiO$_2$ thermal boundary conductance. A relatively wide range of numbers have been reported for this TBC per unit length.[9, 17-19] From their breakdown studies of CNT devices on SiO$_2$, Pop et al[9, 18] extrapolated the TBC to be between 0.12–0.20 WK$^{-1}$m$^{-1}$ for nanotubes between 2–3 nm in diameter. By measuring the temperature distribution in nanotubes of diameter 1.2–2.0 nm with a scanning thermal microscope (SThM), Shi et al[19] obtained TBC values between 0.007–0.06 WK$^{-1}$m$^{-1}$. Finally, from examining the SiO$_2$ substrate coupling of phonon modes at the Brillouin zone center through Raman thermometry, Steiner et al[38] obtained thermal conductance values between 0.03–0.11 WK$^{-1}$m$^{-1}$ for the K, G, and RBM modes of a 1.5 nm diameter CNT. However, given the experimental uncertainties it is challenging to make direct comparison of our simulations with the available data. Nonetheless, our results are well within the experimental range, and in fact help explain why there appear to be a range of values for the TBC with a given substrate, as the variation (and experimental uncertainty) in nanotube diameter naturally leads to variation in the TBC. In addition, the sample-to-sample variability of the CNT-substrate interaction due, for example, to SiO$_2$ surface roughness or adsorbed species in experiments can also give rise to a significant variability in the empirically extracted values for the TBC.

## V. CONCLUSIONS

In summary, we have examined the temperature, diameter, and coupling strength dependence of the thermal boundary conductance (TBC) between carbon nanotubes and amorphous SiO$_2$ using molecular dynamics simulations. We explored the temperature range $T = 200–600$ K, nanotube diameter range $D = 0.81–1.63$ nm, and coupling strength varying by factors $\chi = 1–4$. Our results show the TBC per unit length is sensitive to both CNT diameter and interfacial Van der Waals bond strength. The TBC is found to scale linearly with diameter in armchair CNTs, due to the greater number of phonon modes that can participate in interfacial thermal transport. The TBC scales linearly with the substrate-CNT interaction strength due to the increase in coupling between CNT and substrate phonon modes. The TBC also scales with temperature as ~$T^{1/3}$ in the range considered, independently of the substrate-CNT interaction strength, suggesting that the inelastic scattering of phonons at the interface plays an important role in interfacial thermal transport.

A simple expression for the CNT-SiO$_2$ interface thermal coupling per unit length, which is a good approximation to the MD simulation results can be written as

$$g \approx 0.05 D \chi \left( \frac{T}{200} \right)^{1/3} \qquad (10)$$

where $D$ is in nm and $T$ is in K. With the standard set of parameters $g \approx 0.1$ WK$^{-1}$m$^{-1}$ for a 1.7 nm diameter CNT at room temperature, or approximately $5.8 \times 10^7$ WK$^{-1}$m$^{-2}$ per unit area. The thermal relaxation time of a single CNT on SiO$_2$ is found to be independent of diameter, and approximately 85 ps. We note that such MD simulations yield the lattice contribution to thermal transport, which is thought to be dominant due to the much larger heat capacity of phonons at these temperatures. However, experimental[39] and theoretical[40] work must also examine the role of any electronic degrees of freedom in interfacial thermal transport.

**ACKNOWLEDGEMENT**

This work has been partly supported by the Nanoelectronics Research Initiative (NRI) SWAN center, the NSF CCF-0829907 grant, and a gift from Northrop Grumman Aerospace Systems (NGAS). The molecular images in Fig. 2, 3 and 5 were generated using the graphics program VMD.[41] We acknowledge valuable discussions with Prof. Junichiro Shiomi, and computational support from Reza Toghraee and Prof. Umberto Ravaioli.

**Single-column positioning:**

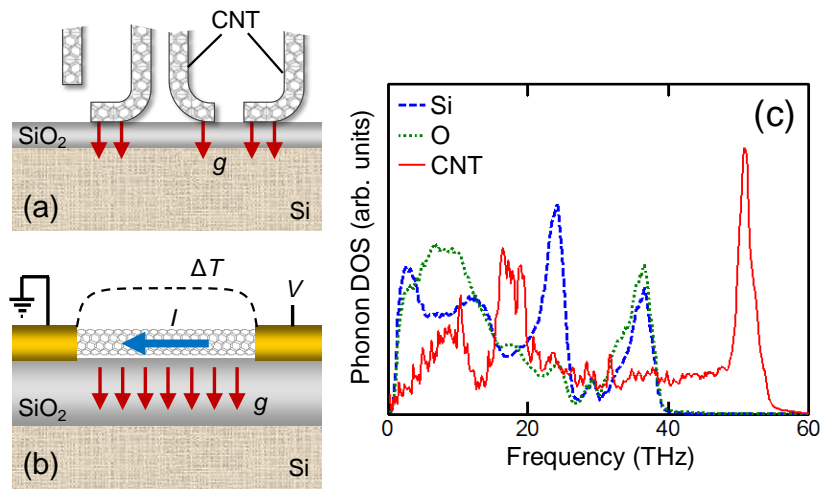

**Figure 1:** (Color online) Schematic of heat dissipation from CNTs to $SiO_2$ in (a) vertical CNT array heat sinks (with native $SiO_2$ layer on Si wafer),[6,7] and in (b) interconnect or transistor applications.[8,9] The red arrows show the direction of heat flow, in particular via the CNT-$SiO_2$ interface. (c) Computed phonon density of states (DOS) in CNT and $SiO_2$ substrate. The CNT phonon DOS ranges from 0–56 THz, whereas that of Si and O atoms range from 0–40 THz.

**Single-column positioning:**

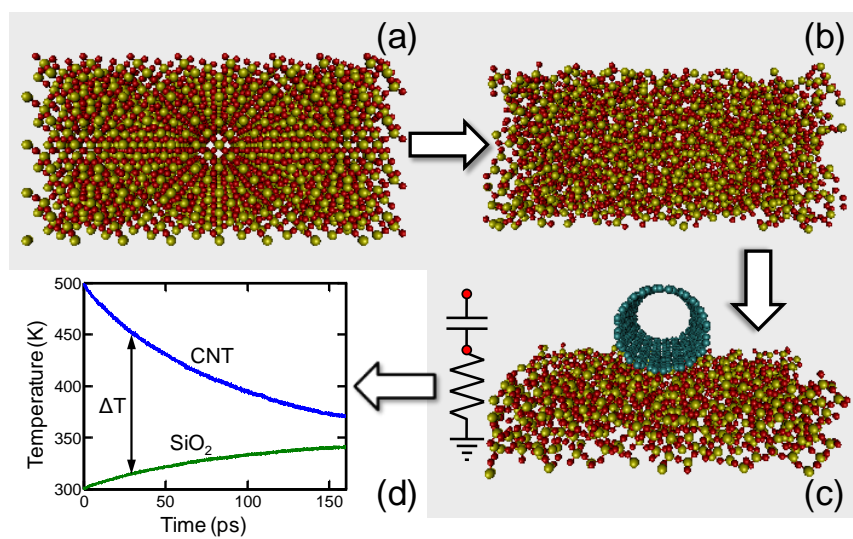

**Figure 2:** (Color online) Generating the simulation structure for a (10,10) nanotube with substrate coupling strength $\chi = 1$. (a) We start with a crystalline $SiO_2$ block and anneal it at 6000 K to produce (b) the amorphous $SiO_2$ domain. (c) We delete the top half of the amorphous block to produce the substrate, then place the CNT on it. Inset shows the thermal circuit formed by the CNT heat capacity and the thermal resistance between CNT and substrate (= $1/g$). (d) Typical simulated temperature transients. The simulation monitors the temperature difference between the CNT and $SiO_2$ as the temperature drop ($\Delta T$) across the interface.

**Single-column positioning:**

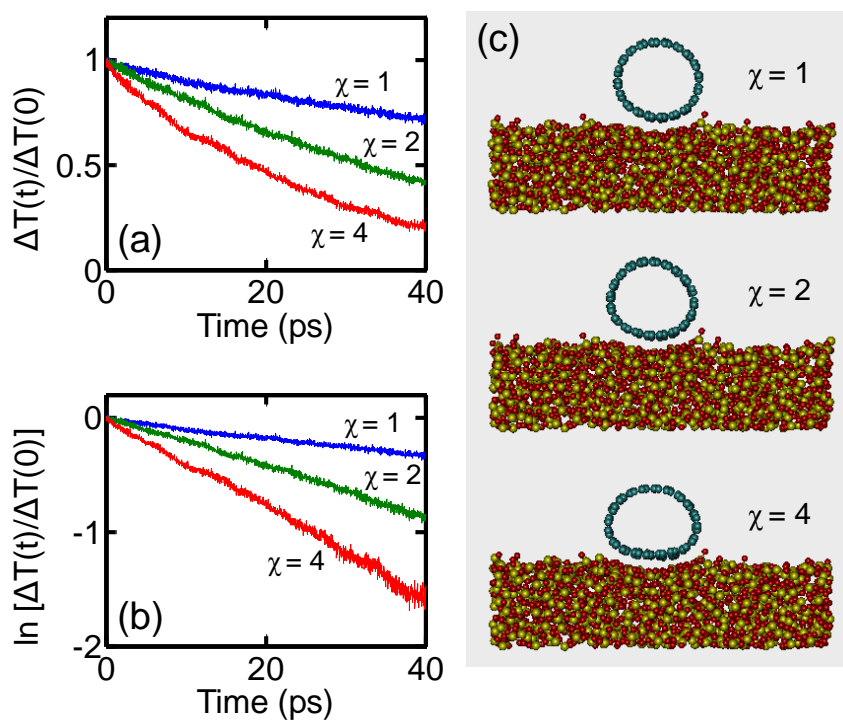

**Figure 3:** (Color online) (a) Decay of the normalized CNT-substrate temperature difference at 300 K averaged over 10 runs, with varying CNT-substrate interaction strength ($\chi$). (b) Natural logarithm of the same plot. The decay of the temperature difference can be fitted to a single exponential decay. The decay time is obtained from the fit gradient in the first 20 ps, and is used to calculate the TBC as described in the text. (c) Cross-sectional profiles of a (10,10) CNT for different values of CNT-substrate coupling strength ($\chi$). As this interaction becomes stronger, the CNT profile is deformed, and its cross-section becomes more elliptical.

**Single-column positioning:**

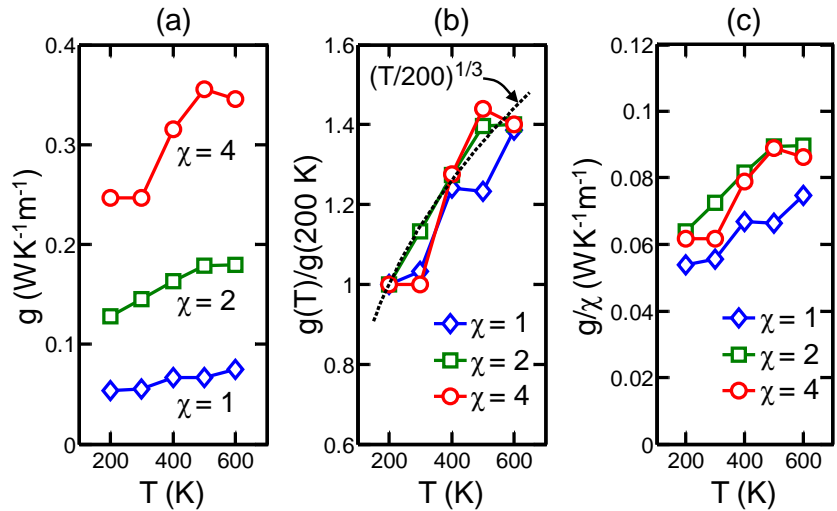

**Figure 4:** (Color online) (a) Computed TBC for different values of $\chi$ from 200–600 K. (b) The TBC normalized with respect to its value at 200 K. Simulations suggest the TBC varies as $\sim T^{1/3}$ for the range of $\chi$ values studied, due to inelastic phonon scattering at the CNT-substrate interface. (c) The TBC normalized with respect to $\chi$. These results suggest there is a nearly linear dependence of the TBC on the strength of the CNT-substrate Van der Waals interaction ($\chi$). This dependence is not captured by conventional theories such as the acoustic or diffuse mismatch models.[35]

**Single-column positioning:**

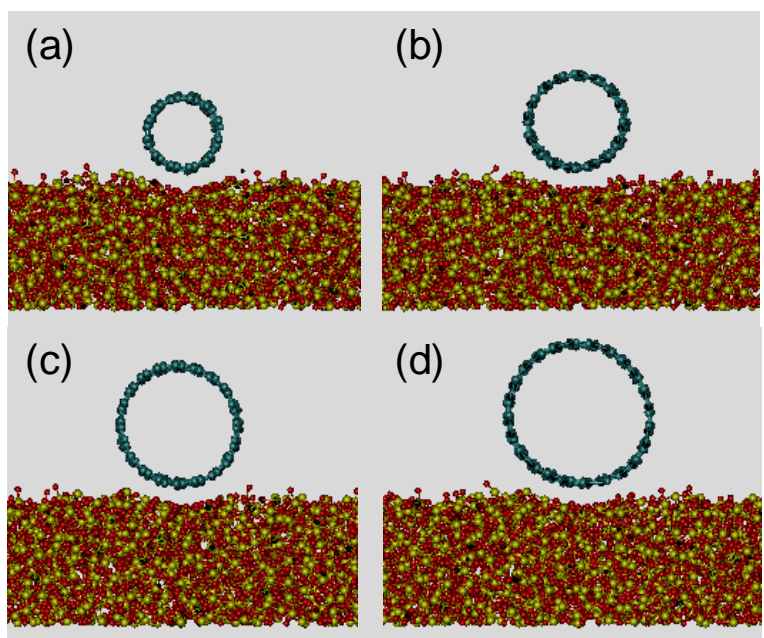

**Figure 5:** (Color online) Cross sectional profiles of (a) (6,6), (b) (8,8), (c) (10,10) and (d) (12,12) CNTs on $SiO_2$, with interaction strength $\chi = 1$. The corresponding diameters are 0.81, 1.08, 1.36 and 1.63 nm, respectively. The CNT and the substrate are equilibrated at 500 K and 300 K, respectively, before their temperature difference is allowed to decay. Averaging over 10 runs, we extract the decay times and use them to compute the TBC values in Fig. 6.

**Single-column positioning:**

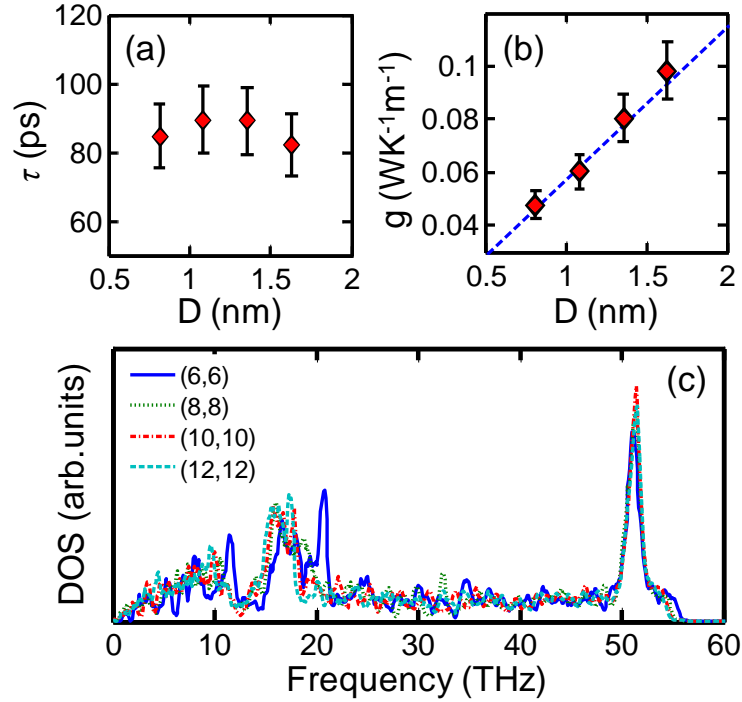

**Figure 6:** (Color online) Diameter dependence of the TBC for armchair CNTs at $T = 300$ K and $\chi = 1$. Error bars represent range from changing VdW potential cut-off.[30] (a) The thermal relaxation time is approximately constant over the range of diameters considered. (b) The calculated TBC is proportional to the CNT diameter (dashed line represents linear fit). This implies that the thermal conductance per unit area normalized by the CNT diameter is constant for the range of diameters studied, $\sim 5.8 \times 10^7$ WK$^{-1}$m$^{-2}$. (c) Normalized phonon density of states for the (6,6), (8,8), (10,10) and (12,12) CNTs. The distribution of the low and high frequency modes is approximately the same for all, suggesting the proportion of low frequency modes stays approximately constant as the CNT diameter increases.